\documentclass[12pt]{article}
\begin{document}
\title{LOOKING BEYOND SPECIAL RELATIVITY}
\author{B.G. Sidharth\\
International Institute for Applicable Mathematics \& Information Sciences\\
Hyderabad (India) \& Udine (Italy)\\
B.M. Birla Science Centre, Adarsh Nagar, Hyderabad - 500 063 (India)}
\date{}
\maketitle
\begin{abstract}
Einstein's Special Theory of Relativity was proposed a little over a hundred years back. It remained a bedrock of twentieth century physics right up to Quantum Field Theory. However, the failure over several decades to provide a unified description of Electromagnetism and Gravitation or alternatively, Quantum Theory and General Relativity has finally lead researchers to abandon the differentiable spacetime manifold on which all of the above was based. In the most recent approaches we consider a spacetime that is discretized or is noncommutative.  This immediately leads to corrections to the Special Theory of Relativity, more specifically to Lorentz Symmetry. It is quite significant that there are observational indicators, particularly in ultra high energy cosmic rays which suggest that such corrections are indeed there. We examine the whole issue in this paper.
\end{abstract}
\section{Introduction}
Just over a hundred years ago, Einstein put forward his Special Theory of Relativity (and subsequently his General Theory of Relativity). Special Relativity has been a bedrock of much of twentieth century physics including Quantum Field Theory. But there are two implicit assumptions.
The first is that spacetime is a differentiable manifold. This has been recognized for a long time. As V.L. Ginzburg puts it [1] 'The special and general relativity theory, non-relativistic quantum mechanics and present theory of quantum fields use the concept of continuous, essentially classical, space and time (a point of spacetime is described by four coordinates $x_\imath = x,y,z,ct$ which may vary continuously). But is this concept valid always? How can we be sure that on a "small scale" time and space do not become quire different, somehow fragmentized, discrete, quantized? This is by no means a novel question, the first to ask it was, apparently Riemann back in 1854 and it has repeatedly been discussed since that time. For instance, Einstein said in his well known lecture "Geometry and Experience" in 1921: "It is true that this proposed physical interpretation of geometry breaks down when applied immediately to spaces of submolecular order of magnitude. But nevertheless, even in questions as to the constitution of elementary particles, it retains part of its significance. For even when it is a question of describing the electrical elementary particles constituting matter, the attempt may still be made to ascribe physical meaning to those field concepts which have been physically defined for the purpose of describing the geometrical behavior of bodies which are large as compared with the molecule. Success alone can decide as to the justification of such an attempt, which postulates physical reality for the fundamental principles of Riemann's geometry outside of the domain of their physical definitions. It might possibly turn out that this extrapolation has no better warrant than the extrapolation of the concept of temperature to parts of a body of molecular order of magnitude."\\
'This lucidly formulated question about the limits of applicability of the Riemannian geometry (that is, in fact macroscopic, or classical, geometric concepts) has not yet been answered. As we move to the field of increasingly high energies and, hence to "closer" collisions between various particles the scale of unexplored space regions becomes smaller. Now we may possibly state that the usual space relationships down to the distance of the order of   are valid or, more exactly, that their application does not lead to inconsistencies. It cannot be ruled out that, the limit is nonexistent but it is much more likely that there exists a fundamental (elementary) length   which restricts the possibilities of classical, spatial description. Moreover, it seems reasonable to assume that the fundamental length   is, at least, not less than the gravitational length $l_g = \sqrt{Gh/c^3} \sim 10^{-33}cm$.\\
'...It is probable that the fundamental length would be a "cut-off" factor which is essential to the current quantum theory: a theory using a fundamental length automatically excludes divergent results.'\\
Einstein himself was aware of this situation. As he observed [2], "...It has been pointed out that the introduction of a spacetime continuum may be considered as contrary to nature in view of the molecular structure of everything which happens on a small scale. It is maintained that perhaps the success of the Heisenberg method points to a purely algebraic method of description of nature that is to the elimination of continuous functions from physics. Then however, we must also give up, by principle the spacetime continuum. It is not unimaginable that human ingenuity will some day find methods which will make it possible to proceed along such a path. At present however, such a program looks like an attempt to breathe in empty space."\\
Attempts were made, starting with the work of Snyder and others in the late 1940s to introduce minimum and fundamental intervals into spacetime [3,4,5] to overcome the divergences encountered in Quantum Electro Dynamics. This would mean that we discard smooth spacetime. The divergences in question themselves arise not from classical considerations but rather from the Quantum Mechanical Heisenberg Uncertainty Principle:
\begin{equation}
\Delta x \cdot \Delta p \geq \hbar\label{eq}
\end{equation}
As can be seen from (\ref{e1}), $\Delta x \to 0$, as would be meaningful in smooth space, would imply the unphysical fact that $\Delta p \to \infty$. Dirac himself noticed this after he formulated his relativistic electron equation [6]. The velocity of the  electron would equal that of light. His explanation was that we have to take averages over the Compton scale to eliminate zitterbewegung effects, to get meaningful physics. It is true that in Quantum Field Theory the divergence problem was circumvented by renormalization, which was in effect giving up attempts to exactly specify certain physical quantities, as Nambu observed [7]. However as Dirac himself noted [8], "I am inclined to suspect that the renormalization theory is something that will not survive in the future, and that the remarkable agreement between its results and experiments should be looked on as a fluke..."\\
More recently the fruitless decades of attempts to unify Gravitation and Electromagnetism or General Relativity and Quantum Mechanics have lead to the realization that we have to discard smooth spacetime. Quantum Gravity approaches, the author's fuzzy spacetime formulation and String Theory work with minimum spacetime intervals [9]. As t' Hooft observed [10], "It is somewhat puzzling to the present author why the lattice structure of space and time had escaped attention from other investigators up till now..." So today the implicit assumptioin of smooth spacetime of twentieth century physics is being reexamined.\\
A second and even less obvious assumption of twentieth century physics has been the acceptance of the existence of absolute or fundamental laws, a consequence of the reductionist approach. This has lead to several paradoxes like that of the arrow of time, the Boltzmann paradox, and so on. At a classical level, the work of Prigogine and his ideas of cooperative behavior and complex systems gave a contra view and could even resolve problems of the arrow of time and the Boltzmann paradox [11,12,13,14]. Indeed, as we will see in a little more detail in the sequel, we should be rather recognizing the universe to be "thermodynamic", in the sense that the laws of physics have a stochastic underpinning [15,5].\\
In a private communication to the author shortly before his death, Prof. Prigogine wrote [14], "I agree with you that spacetime has a stochastic underpinning". These ideas are summarized in the recent book of Robert B. Laughlin where he argues against the reductionist approach [16]. As he puts it, "...I must openly discuss some shocking ideas: The vacuum of spacetime as 'matter', the possibility that relativity is not fundamental, the collective nature of computability, epistemological barriers to theoretical knowledge, similar barriers to experimental falsification and the mythological nature of important parts of modern theoretical physics...". The author himself has been in the forefront of many of these radical new developments and we will touch upon a few which have direct bearing to Special Relativity.
\section{Violations of Lorentz Symmetry}
Given the two new inputs discussed in Section 1, namely a minimum spacetime interval and a non reductioinist approach, it is natural to expect violations of Special Relativity, more specifically Lorentz symmetry. Such violations however would be small and in a suitable limit we should recover Special Relativity.\\
In fact given minimum intervals it is well known that we have, what has now come to be known as a noncommutative geometry. This is given by
$$\left[x,y\right] = \left(\imath a^2 /\hbar\right)L_z, \left[t,x\right] = \left(\imath a^2 / \hbar c\right) M_x, \, etc.$$
\begin{equation}
\left[x,p_x\right] = \imath \hbar \left[1 + \left(a/\hbar \right)^2 p^2_x\right]; etc/\label{e2}
\end{equation}
and similar relations, where   denotes the minimum extension. One interesting feature of (\ref{e2}) is that two independent coordinates $x$ and $y$ no longer commute. This is also characteristic of all modern Quantum Gravity and String Theory approaches.\\
Another interesting feature of (\ref{e2}) is brought out by the last of these equations: It is as if
\begin{equation}
\hbar \to \hbar \left[ 1 + \left(a / \hbar \right)^2 p^2\right]\label{e3}
\end{equation}
Consequently as can be easily shown [17] the energy momentum relation gets modified because of (\ref{e3}) and now becomes
\begin{equation}
E^2 = m^2 + p^2 + \alpha l^2 p^4\label{e4}
\end{equation}
where   is the fundamental length and $\alpha$ is a dimensionless constantof order unity. A consequence of (\ref{e4}) is that the Klein-Gordan and Dirac equations get modified [17,18].\\
In terms of observational effects this would imply a modification of the Compton scattering formula which now becomes
\begin{equation}
k = \frac{mk_0 + \alpha \frac{l^2}{2}\left[Q^2 + 2mQ\right]^2}{\left[m + k_0 \left(1 - cos \Theta \right)\right]}\label{e5}
\end{equation}
where we use natural units $c = \hbar = 1$ or, in terms of frequencies, we would have
\begin{equation}
h\nu = h \nu_0 \left[1 + 0\left(l^2\right)\right]\label{e6}
\end{equation}
As can be seen, both (\ref{e5}) and (\ref{e6}) return to the usual formulae in the limit in which we neglect $l^2$ . Equation (\ref{e6}) means that due to Lorentz symmetry violation given in (\ref{e4}) the frequency is increased or the speed of propagation for a given frequency is increased. It is interesting to note that such effects are detectable, for example in ultra high energy cosmic rays and latest data seems to be indicative of this [19].\\
On the other hand such violations of Lorentz symmetry have been proposed recently by Glashow, Coleman, Carroll, Jacobsen and others, though from a phenomenological point of view [20,21,22,23,24]. The major motivation in this case appears to be the fact that Lorentz symmetry in the context of cosmic rays implies the well known GZK cut off (Cf.ref.[17]). To put it simply the number of cosmic ray events goes down inversely as the third power of the energy, due to the presence of the cosmic microwave background. This in turn implies that no cosmic rays with energy greater than or of the order of $10^{20}eV$ should be observed, given Lorentz symmetry. However a number of such events have been reported (Cf.ref.[17]).\\
We would like to stress that from a fundamental point of view such violations are to be expected due to the non smooth nature of spacetime as pointed out by the author some years ago [5].\\
Wer can come to a similar conclusion from a totally different point of view, one that refers to the non reductionist approach. It has been shown by the author that the photon can be considered to have a non zero but small mass $\sim 10^{-65}gms$ [25,26]. Interestingly from a thermodynamic point of view this mass is the minimum allowable mass in the universe [27]. Using this and the fact that from black hole thermodynamics, the minimum temperature of the universe $\sim 10^{-28}K$, we can deduce using the Langevin equation that the photon travels like a Newtonian particle without external forces with a uniform velocity which is velocity of light (Cf.ref.[25] for details).\\
This in turn implies that for radiation we would have a dispersive group velocity for waves of a given frequency. In fact we would have
\begin{equation}
u_\gamma = c \left[1 - \frac{m^2_\gamma c^4}{\hbar^2 \nu^2}\right]^{1/2}\label{e7}
\end{equation}
(Cf.ref.[26]). As noted above such dispersive cosmic ray events might already have been observed over the past few years [19].\\
We would now argue that interestingly, the violations of the above Lorentz symmetry which we have deduced from two apparently different view points namely a noncommutative geometry and a photon mass, are actually the same. We will give two arguments for this: The first argument is that as was shown in detail [28,29], if we do not neglect terms $\sim l^2$ in gauge theory, then the gauge field $A_\mu$ (Abelian or non Abelian) gets a new term:
\begin{equation}
A_\mu \to A_\mu + \partial_\mu f\label{e8}
\end{equation}
The last terms in (\ref{e8}) represents the $\sim l^2$ effect and plays the role of the Higgs Field in the usual theory (Cf.ref.[9,30]), so that the kinetic energy term is given by
\begin{equation}
\left| D_\mu f\right|^2 = \beta \left|A_\mu \right|^2 \left|f\right|^2 \cdots\label{e9}
\end{equation}
where $\beta$ is a suitable constant. One can immediately see from (\ref{e9}) that it defines the mass term, exactly as with the Higgs symmetry breaking mechanism, viz., 
\begin{equation}
m^2 A_\mu A^\mu\label{e10}
\end{equation}
The original problem with gauge fields was that the mass term given in (\ref{e10}) could not be included in the Lagrangian, as it was not gauge invariant. It now finds its way back into the Lagrangian via the invariant kinetic energy term (\ref{e9}) thanks to the additional $\sim l^2$ terms that are now taken into account, and which play the same role as the Higgs field with the symmetry breaking mechanism (Cf.ref.[28,29] for a detailed description). The point now is that in the above considerations  could very well stand for the electromagnetic field itself. In the usual point spacetime formulation however in which $\sim l^2$ terms had to be neglected, the contribution of the gauge term $f$ would vanish.\\
We could give an alternative derivation. The noncommutative geometry in (\ref{e2}) above necessitated by minimum spacetime intervals, as is well known leads in all these approaches to
\begin{equation}
\Delta x = \frac{\hbar}{\Delta p} + \frac{l^2}{\hbar} \Delta p\label{e11}
\end{equation}
The relation (\ref{e11}) which replaces (\ref{e1}) implies
$$\Delta x^2 = \frac{\hbar^2}{\Delta p^2} + \frac{l^4}{\hbar^2} \Delta p^2 + 2l^2$$
so that
$$\Delta p \to \, \mbox{implies}\, \langle \Delta x^2\rangle_{max} = \frac{\hbar^2}{\Delta p^2} = R^2$$
In the above, $R$ represents the maximum possible extent in the universe namely the radius of the universe $\sim 10^{28}cms$. This yields the limiting mass
$$m_\gamma = \hbar / Rc = 10^{-65}gm$$
which is the previously deduced mass of the photon. We will briefly return to the above considerations.
\section{Issues Relating to Instantaneous Action At A Distance}
We have remarked earlier that there are unphysical zitterbewegung effects within the Compton wavelength. Specifically it is known that if
\begin{equation}
0 < (\vec{r_1} - \vec{r_2})^2 - c^2 (t_1 - t_2)^2 \leq \left(\frac{\hbar^2}{mc}\right)^2\label{e12}
\end{equation}
holds then the particle has a non zero probability of being at points $\vec{r_1}$ and $\vec{r_2}$ even if this interval is space like [31]. It is interesting to note that if in (\ref{e12}) we consider a photon and use its mass then the distance between   and   can be the radius of the universe, even if $t_1 = t_2$. We could interpret this result by saying that the photon could be found instantaneously at two well separated points. Such instantaneous action at a distance which was considered by Feynman and Wheeler in the 1940s [32] has received attention more recently, notably through the work of Hoyle and Narlikar, Chubykalo, Smirnov-Rueda, the author himself and others [33,34,35,36,37]. Before proceeding, let us briefly touch upon the main points of this formulation.\\
We begin with classical electrodynamics. From a classical point of view a charge that is accelerating radiates energy which dampens its motion. This is given by the well known Maxwell-Lorentz equation, which in units $c = 1$, [34], and  being the proper time, while $\imath = 1,2,3,4,$ is (Cf. [34]),
\begin{equation}
m\frac{d^2x^\imath}{d\tau^2} = eF^{\imath k} \frac{dx^k}{d\tau} + \frac{4c}{3} g_{\imath k} \left(\frac{d^3x^\imath}{d\tau^3} \frac{dx^\imath}{d\tau} - \frac{d^3 x^l}{d\tau^3} \frac{dx^\imath}{d\tau}\right)\frac{dx^k}{d\tau},\label{e13}
\end{equation}
The first term on the right side is the usual external field while the second term is the damping field which is added ad hoc by the requirement of the energy loss due to radiation. In 1938 Dirac introduced instead 
\begin{equation}
m\frac{d^2x^\imath}{d\tau^2} = e\left\{F^\imath_k + R^\imath_k\right\} \frac{dx^k}{d\tau}\label{e14}
\end{equation}
where
\begin{equation}
R^\imath_k \equiv \frac{1}{2} \left\{F^{ret_\imath}_k - F^{adv_\imath}_k \right\}\label{e15}
\end{equation}
In (\ref{e15}), $F^{ret}$ denotes the retarded field and $F^{adv}$ the advanced field. While the former is the causal field where the influence of a charge at  $A$ is felt by a charge at $B$ at a distance $r$ after a time $t = \frac{r}{c}$, the latter is the advanced acausal field which acts on $A$ from a future time. In effect what Dirac showed was that the radiation damping term in (\ref{e13}) or (\ref{e14}) is given by (\ref{e15}) in which an antisymmetric difference of the advanced and retarded fields is taken, which of course seemingly goes against causality as the advanced field acts from the future backwards in time. It must be mentioned that Dirac's prescription lead to the so called runaway solutions, with the electron acquiring larger and larger velocities in the absence of an external force. This he related to the infinite self energy of the point electron.\\
As far as the breakdown of causality is concerned, this takes place within a period $\sim \tau$, the Compton time. It was at this stage that Wheeler and Feynman reformulated the above action at a distance formalism in terms of what has been called their Absorber Theory. In their formulation, the field that a charge would experience because of its action at a distance on the other charges of the universe, which in turn would act back on the original charge is given by
\begin{equation}
Re = \frac{2e^2 d}{3 dt} (\ddot{x})\label{e16}
\end{equation}
The interesting point is that instead of considering the above force in (\ref{e16}) at the charge $e$ , if we consider the responses in its neighborhood, in fact a neighborhood at the Compton scale, as was argued recently by the author [37], the field would be precisely the Dirac field given in (\ref{e14}) and (\ref{e15}). The net force emanating from the charge is thus given by
\begin{equation}
F^{ret} = \frac{1}{2} \left\{F^{ret} + F^{adv}\right\} + \frac{1}{2}\left\{F^{ret} - F^{adv}\right\}\label{e17}
\end{equation}
which is the acceptable causal retarded field. The causal field now consists of the time symmetric field of the charge   together with the Dirac field, that is the second term in (\ref{e17}), which represents the response of the rest of the charges. Interestingly in this formulation we have used a time symmetric field, viz., the first term of (\ref{e17}) to recover the retarded field with the correct arrow of time. There are two important inputs which we can see in the above formulation. The first is the action of the rest of the universe at a given charge and the other is minimum spacetime intervals which are of the order of the Compton scale.\\
Feynman and Wheeler stressed that the universe has to be a perfect absorber or to put it simply, every charged particle in the universe should respond back to the action on it by the given charge - and the other is, as stressed by the author, minimum spacetime intervals which are of the order of the Compton scale. We now follow up this line of reasoning, which was essentially from electrodynamics to a characterization of gravitation itself.\\ 
In fact the work done on a charge $e$ at $0$ by the charge $P$ a distance $r$ away in causing a displacement $dx$ is given by (ignoring a cosine which merely gives a small numerical factor),
\begin{equation}
\frac{e^2}{r^2} dx\label{e18}
\end{equation}
Now the number of particles at distance $r$ from $0$ is given by
\begin{equation}
n(r) = \rho (r) \cdot 4\pi r^2\label{e19}
\end{equation}
where $\rho(r)$ is the density of particles. So using (\ref{e19}) in (\ref{e18}) the total work is given by
\begin{equation}
E = \int \int \frac{e^2}{r^2} 4\pi r^2 \rho dxdr\label{e20}
\end{equation}
which can be shown using a uniform average density $\rho$, to be $\sim mc^2$. We thus recover in (\ref{e20}) the inertial energy of the particle in terms of its electromagnetic interactions with the rest of the universe in an action at a distance scheme.\\
Interestingly this can also be deduced in the context of gravitation itself. The work done on a particle of mass   which we take to be a pion, a typial elementary particle, by the rest of the particles (pions) in the universe is given by
\begin{equation}
\frac{Gm^2 N}{R}\label{e21}
\end{equation}
It is known that in (\ref{e21}) $N \sim 10^{80}$ while $R \sim \sqrt{N}l$, the well known Weyl-Eddington formula. Whence the gravitational energy of the pion is given by
\begin{equation}
\frac{Gm^2 \sqrt{N}}{l} = \frac{e^2}{l} \sim mc^2\label{e22}
\end{equation}
where in (\ref{e22}) we have used the fact that
\begin{equation}
Gm^2 \sim \frac{e^2}{\sqrt{N}}\label{e23}
\end{equation}
It must be mentioned that though the Eddington formula and (\ref{e23}) were  originally empirical, they can in fact be deduced from theory in the author's formulation (Cf. for references and a summary [5]).\\
It may also be pointed out that what is clear in the above formulation is the Machian non reductioinist aspect as was remarked in the introduction. We can also look at the above arguments from a slightly different point of view, one which considers the neutral atoms in the universe, which nevertheless have a dipole effect.\\ 
In fact as is well known from elementary electrostatics the potential energy at a distance $r$ due to the dipole is given by
\begin{equation}
\phi = \frac{\mu}{r^2}\label{e24}
\end{equation}
where $\mu = eL, L \sim 10^{-8}cm \sim 10^3 l \equiv \omega l, e$ being the electric charge of the electron for simplicity and $l$ being the electron Compton wavelength. (There is a factor $cos \Theta$ with $\mu$, but on an integration over all directions, this becomes as before an irrelevant constant factor $4\pi$.)\\
Due to (\ref{e24}), the potential energy of a proton $p$ (which approximates an atom in terms of mass) at a distance $r$ (much greater than $L$) is given by
\begin{equation}
\frac{e^2 L}{r^2}\label{e25}
\end{equation}
As there are $N \sim 10^{80}$ atoms in the universe, the net potential energy of a proton due to all the dipoles is given by
\begin{equation}
\frac{Ne^2 L}{r^2}\label{e26}
\end{equation}
In (\ref{e26}) we use the fact that the predominant effect comes from the distant atoms which are at a distance $\sim r$, the radius of the universe.\\
We use again the well known Eddington formula encountered earlier,
\begin{equation}
r \sim \sqrt{N}l\label{e27}
\end{equation}
If we introduce (\ref{e27}) in (\ref{e26}) we get, as the energy $E$ of the proton under consideration
\begin{equation}
E = \frac{\sqrt{N}e^2 \omega}{r}\label{e28}
\end{equation}
Let us now consider the gravitational potential energy $E'$ of the proton $p$ due to all the other $N$ atoms in the universe. This is given by
\begin{equation}
E' = \frac{GMm}{r}\label{e29}
\end{equation}
where $m$ is the proton mass and $M$ is the mass of the universe.\\
Comparing (\ref{e28}) and (\ref{e29}), not only is $E$ equal to $E'$, but remembering that $M = Nm$, we deduce in this fine tuned model, the relation (\ref{e23}), viz.,
$$\frac{e^2}{Gm^2} = \frac{1}{\sqrt{N}}.$$
Furthermore, as we have deduced (\ref{e23}), in effect, in this model we have got a unified description of electromagnetism and gravitation.\\
The points to be stressed are that firstly, we are able to use Quantum Mechanical effects to reconcile action at a distance electrodynamics with Special Relativity, and secondly this is possible because of a Machian or holistic effect, brought out for example by the Feynman perfect absorber condition, or what may be called a fine tuned universe.
\section{Discussion}
As noted in the introduction, Einstein (and much of twentieth century physics) operated in a differentiable spacetime manifold and adopted a reductionist approach. In fact Einstein argued for local realism [38,39]. An element of physical reality corresponding  to a physical attribute of some system is the value that can be predicted with certainty and without disturbing the physical system itself. The concept of locality dictates that we can deal in a physically meaningful manner with parts of the external world without the necessity of dealing with all or any other part of it. So, effectively the elements of physical reality of a given system do not depend on measurements performed on a spatially separated system which is not in direct causal interaction with this system. Causality in this post relativity context, would mean the following: Any two events separated by a space like interval are not in causally connected.\\ 
This seemed counter to Quantum Mechanics. Einstein with Podolsky and Rosen constructed the famous EPR experiment which would prove that it would be possible to obtain information about the angular momentum of a particle A by making a measurement of a distant particle B. In effect information about particle A has been obtained instantaneously without a separate measurement, as would be required in Quantum Theory though not in Classical Theory. Schrodinger argued that there was no contradiction and that this was due to what was then called the non separability property which distinguished Quantum Theory from Classical Theory-the particles A and B, once they interact would form what may be called an entangled system in modern terminology: they would be described by a single wave function, requiring a single measurement. Several experiments starting with that of Aspect in the 1980s have since vindicated this property and today Quantum entanglement is at the heart of the emerging subject of Quantum Computing. Nevertheless it has been argued time and again that all this does not contradict Einstein's causality, that is there is no superluminal transfer of information. As this has been extensively discussed in the literature, we merely cite some references. [40,41].

\end{document}